\documentclass[12pt,preprint]{aastex}
\usepackage{natbib}
\shorttitle{Statistics of Magnetic Fields for OB Star}
\shortauthors{{A. F. Kholtygin $^{1,}$\footnote{*E-mail: afkholtygin@gmail.com}, 
    S. N. Fabrika$^{2}$, N. A. Drake$^{1}$, V. D. Bychkov$^{2}$, L. V. Bychkova$^{2}$,  
   G. A. Chountonov$^{2}$,  T. E. Burlakova$^{2}$, and G. G. Valyavin$^{3}$}}

\usepackage{graphicx}
\usepackage{amssymb} 
\usepackage{amsmath} 

\newcommand{\bc}{\begin{center}}
\newcommand{\ec}{\end{center}}
\newcommand{\be}{\begin{equation}}
\newcommand{\ee}{\end{equation}}

\newcommand{\ov}{\overline}
\newcommand{\bea}{\begin{eqnarray}}
\newcommand{\eea}{\end{eqnarray}}

\newcommand{\ds}{\displaystyle}

\newcommand{\lem}{\leqslant}

\newcommand{\sz}{\scriptsize}

\renewcommand{\abstractname}{}
\def\bitem#1#2\par{\footnotesize\noindent\hangindent1.5\parindent\hangafter=1
7\em#1\rm#2\par\medskip}
\def\bbitem#1#2\par{\footnotesize\noindent\hangindent1.5\parindent
\hangafter=1\em#1\rm#2\par\medskip}

\begin{document}
\thispagestyle{empty}

\bc
{\large\bf Statistics of Magnetic Fields for OB Stars
}
\ec

\bc
{\sf
    A. F. Kholtygin $^{1,}$\footnote{*E-mail: afkholtygin@gmail.com}, 
    S. N. Fabrika$^{2}$, N. A. Drake$^{1}$, V. D. Bychkov$^{2}$, L. V. Bychkova$^{2}$,  
    G. A. Chountonov$^{2}$,  T. E. Burlakova$^{2}$, and G. G. Valyavin$^{3}$
}
\ec

{\small
 \noindent
    $^{1}$Astronomical Institute, St. Petersburg State University, Bibliotechnaya pl. 2, 
          Petrodvorets, 198504 Russia\\
    $^{2}$Special Astrophysical Observatory, Russian Academy of Sciences, Nizhnii Arkhyz,
                Karachai-Cherkessian Republic, 357147 Russia\\
        $^{3}$National Astronomical Observatory of Mexico, Mexico
                Received July 13, 2009\\ 
}
\begin{abstract}
$\bf Abstract$ -- Based on an analysis of the catalog of magnetic fields, we have investigated the statistical
properties of the mean magnetic fields for OB stars. We show that the mean effective magnetic field ${\cal B}$
of a star can be used as a statistically significant characteristic of its magnetic field. No correlation has been
found between the mean magnetic field strength ${\cal B}$ and projected rotational velocity of OB stars, which is
consistent with the hypothesis about a fossil origin of the magnetic field. We have constructed the magnetic
field distribution function for B stars, $F({\cal B})$, that has a power-law dependence on ${\cal B}$ with an exponent of $\approx -1.82$. We have found a sharp decrease in the function $F({\cal B})$F for ${\cal B}\lem 400~G$ that may be related to rapid dissipation of weak stellar surface magnetic fields.
\end{abstract}

\section{INTRODUCTION}
\label{s.Intro}
A sizeable fraction of stars possess magnetic fields accessible to present-day measurements. The magnetic fields 
of stars may affect significantly their evolution. F- and later-type stars with masses $M<1.5-2\,M_{\odot}$ often
possess strong and irregular magnetic fields whose generation is associated with the action of a dynamo mechanism,
which is eventually reduced to the conversion of part of the mechanical energy of stellar rotation into the energy of the generated magnetic field. More massive early-type stars with $M>1.5-2\,M_{\odot}$ without convective envelopes are characterized by the presence of regular magnetic fields whose origin is still not quite clear.

A number of researchers believe that the dynamo mechanism is also efficient for hot OBA stars. In this case, it is assumed that the magnetic field is generated in the convective stellar core and individual magnetic flux tubes then rise through the radiative stellar envelope \cite{2003ApJ...586..480M}. These authors hypothesize that a stellar surface magnetic field with a strength of several hundred gauss is generated in this way, but it remains unclear how a regular field structure can be formed through the random process of the buoyant rise of magnetic flux tubes.

The hypothesis about a {\it fossil origin} of the magnetic field in OBA stars is currently more popular. It suggests that the stellar magnetic field observed at the present epoch is a (fossil) remnant of the magnetic field of the molecular cloud in which the star was formed \cite{2006A&A...450.1077B}.

Statistical studies of the correlation between basic stellar characteristics and magnetic field strengths can shed light on the generation mechanisms and evolution of the magnetic fields in OB stars. The recently published catalog by \cite{2009MNRAS.394.1338B} provides information about magnetic field measurements for more than 1000 OBA stars. We used the data from this catalog in combination with the data obtained after the catalog was submitted for publication for our investigation.

In this paper aimed at investigating the statistical properties of the magnetic fields for an ensemble of hot stars, we consider the observational data on the magnetic fields of OB stars. The statistical properties of an ensemble of magnetic fields in A stars will be investigated in our next publications.

Here, we briefly describe the currently used methods of magnetic field measurements and describe the choice of magnetic field characteristics to be used for a statistical analysis. Next, we present the catalog of magnetic fields for OBA stars and the results of recent measurements for O and B stars. We analyze the statistical properties of the magnetic fields for OB stars. The conclusions are presented in the concluding section of the paper.

\section{STATISTICAL DESCRIPTION OF MAGNETIC FIELDS}
One of the main methods for determining the magnetic field strength is to measure the Zeeman shift $\Delta \lambda$ between the left (L) and right (R) hand circularly polarized components of the line profile \cite{1947ApJ...105..105B}. The shift is proportional to the longitudinal magnetic field component $B_l$ averaged over the stellar disk, which is often called an {\it effective magnetic field}.

In recent years, to measure $B_l$, a statistically significant signal has been sought for in the line profiles directly in the Stokes V parameter. The LSD (least-squares deconvolution) technique \cite{1997MNRAS.291..658D},is used to determine the Stokes V parameter averaged over a large number of lines.
The effective magnetic field depends on the stellar rotation phase and varies over a wide range from a minimum value of $B_{min}$ to a maximum value of $B_{max}$, with $B_{min}$ and $B_{max}$ often having opposite signs. This means that $B_l$ is unsuitable for statistical studies of the magnetic fields for a large ensemble of stars. For this reason, a global field characteristic that can be obtained from observations and that is not subject to large variations depending on the rotation phase $\phi$ at which the field measurements were made should be used. Since the largest variations in $B_l$ with rotation phase are characteristic of a dipole global stellar magnetic field and the contribution from higher-order harmonics (quadrupole, octupole, and higher-order harmonics) for the B stars we consider is relatively small (see, e.g., \cite{2008MNRAS.386.1947B}), it will suffice to analyze a dipole field.

In the model of a rotating magnetic dipole, $B_l$ depends on the stellar rotation phase $\phi$ (\cite{1967ApJ...150..547P})
\be
\label{Eq.Bl-rot}
{B}_l= {B}_l(\phi) = 
B_p \,\frac{15+u}{20(3-u)} \left[ \cos \beta \cos i
+ \sin\beta \sin i \cos 2 \pi (\phi -\phi_0) \right]\, ,
\ee
where $B_p$ is the polar field strength, $\beta$ is the angle between the magnetic dipole axis and the rotation
axis, $i$ is the inclination of the rotation axis to the line of sight, $\phi_0$ is the rotation phase at which ${B}_l(\phi)$
is at a maximum, and $u$ is the limb-darkening coefficient. For O and B stars, $u=0.35$~(\cite{1997A&A...317..723T}) can be used.

The phase-averaged ratios $\ov{B_l}/B_p$, where 
$
\ds  \ov{B}_l = \!\int_{0}^{2\pi}\!\! {B_l}(\phi)\, d\phi \, ,
$
for all of the possible $\beta$ and $i$ do not exceed 0.3. $\ov{B}_l$ depends significantly on the inclination $i$, which is determined by the random orientation of the stellar rotation axis, and on the angle $\beta$, which also varies over a wide range. Thus, the mean field strength $\ov{B}_l$ cannot be used for statistical studies. In addition, the times at which the field measurements were made are sometimes distributed very irregularly in stellar rotation phase.

As one of the possible characteristics of the mean magnetic field that depends weakly on the random values of $i$ and $\beta$, we will consider the root-meansquare (rms) field widely used in the literature (see, e.g., \cite{1993A&A...269..355B}),
\be
\label{Eq.Baver}
{\cal B} = 
\sqrt{\frac{1}{n}\sum_{i=1}^{n} ({B}_l^{i})^2}
\, .
\ee
Here, the summation is over all $n$ field measurements. In addition to this quantity, we will introduce another field characteristic,
\be
\label{Eq.BaverMinMax}
{\cal B}_{\mbox{\sz m}} = 
\sqrt{\frac{1}{2}(B_{\mbox{\sz min}}^2 + B_{\mbox{\sz max}}^2)}\, ,
\ee
where $B_{\mbox{\sz min}}$ and $B_{\mbox{\sz max}}$ are the minimum (given the sign) and maximum values of ${B_l}$ determined 
from all field measurements.

\cite{2008AstBu..63..139R} suggested the extreme field determined from all measurements as a quantity that characterizes the mean stellar magnetic field:
\be
\label{Eq.Bextr}
{\cal B}_{\mbox{\sz extr}} = 
\max\limits_{i=1,\dots, n} 
 \left( |{B}_l^{i}| \right)  \, .
\ee
The following quantity is commonly used to characterize the accuracy of a field measurement:
\be
\label{Eq.Baver_sigma}
\sigma_{\cal B} = 
\sqrt{\frac{1}{n}\sum_{i=1}^{n} 
\left(\sigma_{{B}_l^{i}}\right)^2}
\, .
\ee
where $\sigma_{{B}_l^{i}}$ is the rms error of the $i$th field measurement. Although the quantity $\sigma_{\cal B}$ widely used in the literature is not the standard deviation of the rms field $\cal B$, since the $\cal B$ itself is not a normally distributed random variable, it can serve as a measure of the accuracy of the entire series of field measurements. It is generally believed that if ${\cal B} > 2\, \sigma_{\cal B}$, then the measured field strengths are real. The standard $\chi^2$ statistics defined as
\be
\label{Eq.Baver_chi2}
\chi^2 = \sum_{i=1}^{n} \frac{\left({B}_l^{i}\right)^2}
         {\left(\sigma_{{B}_l^{i}}\right)^2} \, .
\ee
is used to estimate the extent to which the field measurements for a specific star are reliable. The reduced $\chi^2/n$ value is commonly used instead of $\chi^2$ (see, e.g., \cite{2009MNRAS.394.1338B}). When the star has no magnetic field, the expectation values 
of ${B}_l^{i}$ are zero. In this case, the smallness of $\chi^2/n$ compared to unity suggests that the hypothesis of ${B}_l^{i}=0$
is valid and $\chi^2/n\gg 1$ suggests that the field measurements are real.

The estimates of $\sigma_{{B}_l^{i}}$ can be very uncertain and change by an order of magnitude or more for the same
object during one set of observations (\cite{2007MNRAS.382.1690L}). Therefore, whether the $\chi^2$ test is applicable for estimating the extent to which the field measurements are reliable remains an open question.

To ascertain the extent to which the quantities $\ds {\cal B}$, $\ds {\cal B}_{\mbox{\sz m}}$ and $\ds {\cal B}_{\mbox{\sz extr}}$
introduced above can be used as characteristics of the mean magnetic field, let us calculate their values in the limit of an infinite number of measurements ($n \to \infty$) uniformly distributed in stellar rotation phase. Using Eq.~(1) and replacing the summation with integration over the phases $\phi$, we will obtain
\be
\label{Eq.BaverPhase}
{\cal B}  
\xrightarrow[{n\to\infty}]{}
 B_p\left[\frac{1-u/15}{4(1-u/3)}
\sqrt{\cos^2 \beta \cos^2 i + \frac{1}{2}\sin^2\beta \sin^2 i}\right] \, .
\ee
Similarly, we can also find that
\be
\label{Eq.BaverPhaseMinMax}
{\cal B}_{\mbox{\sz m}} 
\xrightarrow[{n\to\infty}]{} 
 B_p \left[\frac{1-u/15}{4(1-u/3)}
\sqrt{\cos^2 \beta \cos^2 i + \sin^2\beta \sin^2 i} \right] \, .
\ee
and
\be
\label{Eq.BaverPhaseExtr}
{\cal B}_{\mbox{\sz extr}} \xrightarrow[{n\to\infty}]{} 
 B_p\left\{\dfrac{1-u/15}{4(1-u/3)}
\max \left[ \cos(\beta+i), \cos(\beta-i) \right]\right\} \, .
\ee
In Fig. 1, the rotation-phase-averaged ratios ${\cal B}/B_p$, ${\cal B}_{\mbox{\sz m}}/B_p$ and ${\cal B}_{\mbox{\sz extr}}/B_p$ are plotted against the inclination of the rotation axis $i$ for angles $\beta$ equal to $30^{\circ}$, $45^{\circ}$ and $60^{\circ}$. We see that these dimensionless ratios vary within the range 0.12~--~0.30 with mean values close to 0.20.

In real observations, the number of magnetic field measurements is usually small. To understand how the values of the quantities we consider depend in this case on the number of field measurements for various angles $i$ and $\beta$, let us model the field measurement process in the following way. Suppose that the field measurements for each of the possible angles $i$ and $\beta$ were performed $N_{\mbox{\sz mf}}$ times, where the number $N_{\mbox{\sz mf}}$ can be equal to one.

We will assume that the stellar rotation phase $\phi$ at which the field was measured is a random variable uniformly distributed in the interval $[0,1]$. Suppose also that $i$ is a random variable determined by the random orientations of the stellar rotation axis. At the same time, the angle $\beta$ probably varies over a narrower range ($\pm \sim 15^{\circ}$ ) with a mean value close to $45^{\circ}$ (see, e.g., \cite{2007AN....328.1170K}; \cite{2007A&A...475.1053A}). For this reason, we performed our calculations for two ranges of $\beta$: $30-60^{\circ}$ and $0-90^{\circ}$.

The calculations were performed as follows: we chose $\approx 5000$ random values of the angles $i$ and $\beta$ that
varied within the above ranges. The number $N_{\mbox{\sz mf}}$ of random rotation phases $\phi$ was determined for each pair of $i$ and $\beta$. $B_l$ was determined for each of these phases $\phi$ from Eq.~(1). The values of $B_l$ obtained were used to calculate the ratios ${\cal B}/B_p$, ${\cal B}_{m}/B_p$ and ${\cal B}_{\mbox{\sz extr}}/B_p$ from Eqs. (2)~--~(4). Since these ratios do not depend on $B_p$, the latter was taken to be equal to one. For ${\cal B}/B_p$, ${\cal B}_{m}/B_p$ and ${\cal B}_{\mbox{\sz extr}}/B_p$, we determined the mean values of these quantities and the corresponding standard deviations $\sigma$, $\sigma_m$ and $\sigma_{\mbox{\sz extr}}$ in a standard way. The results of our numerical experiment are presented in Fig.~2.

Analysis of Fig.~2 shows that both ${\cal B}$ and ${\cal B}_{m}$ vary within a narrow range from $\approx 0.17$ to $\approx 0.20$ with a median value of $\approx 0.19$. The mean values are ${\cal B}_{\mbox{\sz extr}} \approx 0.23$. Note that $\sigma$, $\sigma_m$, and $\sigma_{\mbox{\sz extr}}$ are relatively small even for $N_{\mbox{\sz mf}}= 2$ and the characteristics we consider are statistically significant for any angles $\beta$.

The differences between the mean ${\cal B}$ and ${\cal B}_{m}$ are statistically insignificant. Below, we will use ${\cal B}$, because it is this quantity that is usually provided in the papers devoted to magnetic field measurements.
Since $\sigma$, $\sigma_m$ and $\sigma_{\mbox{\sz extr}}$ are close to the field strengths for $N_{\mbox{\sz mf}}=1$ (in this case, ${\cal B}_{\mbox{\sz extr}}=|B_{\mbox{\sz min}}|=B_{\mbox{\sz max}}=B_l$), we will exclude single measurements from our statistical analysis.

\section{OBSERVATIONAL DATA}
\label{ss.MeasureResults}
In many papers, the statistical properties of the magnetic fields for O and B stars are investigated by analyzing the data from the catalog by \cite{2009MNRAS.394.1338B}. This catalog contains the published data accessible to the authors along with the results of unpublished observations, both their own observations and those of other authors. The catalog provides information about the magnetic field measurements for 1212 main-sequence and giant stars, 610 of which are chemically peculiar (CP) stars. The relationship between atmospheric chemical anomalies and stellar magnetization has long been known (see, e.g., \cite{1974ARA&A..12..257P}). The number of CP stars with respect to the normal ones has been variously estimated to be no more than 15$\%$ (\cite{1974ARA&A..12..257P}). Nevertheless, despite their relatively small number, they have been studied more extensively with regard to magnetic field measurements. In this paper, we will consider a sample of OB stars from the catalog by \cite{2009MNRAS.394.1338B}. The latter provides information about the magnetic field measurements performed for 15 O stars and 416 B stars, 239 of which are CP stars.

New magnetic field measurements for OB stars have appeared since the catalog by \cite{2009MNRAS.394.1338B} was submitted for publication (\cite{2008A&A...483..857S}; \cite{2008MNRAS.387L..23P}; \cite{2007AN....328.1170K}; \cite{2008MNRAS.389...75B}; \cite{2008ApJ...686.1269M}; \cite{2009MNRAS.398.1505S}). Below, we use the new magnetic field measurements for OB stars that are presented in the cited papers and that were not included in the catalog by \cite{2009MNRAS.394.1338B} to analyze the statistical properties of the magnetic fields for OB stars.

Figure 3 presents the spectral-type-averaged magnetic fields of OB stars. Since the number of magnetic field measurements for O stars is small (for eight stars), the figure shows the mean value of $\cal B$ for all O stars with measured magnetic fields. The number of stars of different spectral subtypes with measured magnetic fields is also indicated in the figure. The sharp increase in the number of stars with measured fields for the spectral subtypes B8 and B9, along with the overall increase in the number of stars of these spectral types, can also be explained by the appearance of CP stars with helium abundance anomalies in the corresponding temperature range.

Although the interval of $\cal B$ for stars of the same subtype is very large and its standard deviations for the spectral subtypes B1--B3 and B5--B9 are comparable to the values of $\cal B$ themselves, we can reach a tentative conclusion about the existence of a jump in the above mean values when passing from O stars to B stars. For B4 stars, the catalog provides statistically significant magnetic field measurements only for one object (HD 175362) and, hence, the high value of ${\cal B}=3738$~G presented in Fig. 3 can be a random spike.

The cause of such a jump in $\cal B$ still remains unclear. If the magnetic flux for OB stars is assumed to be approximately constant (see, e.g., \cite{2008MNRAS.387L..23P}), then this jump could be partly explained by larger radii of O stars than those of B stars. This effect may also be related to a significant mass loss rate by O stars and the removal of magnetic flux by a stellar wind.

\section{STATISTICAL PROPERTIES OF THE MAGNETIC FIELDS FOR OB STARS}
\label{s.MFstatOBstars}

To ascertain how the mean stellar magnetic fields depend on stellar parameters, we used ${\cal B}$ from the catalog by \cite{2009MNRAS.394.1338B} and the new values presented in the papers cited above. The catalog by \cite{2009MNRAS.394.1338B} provides {\it all} magnetic field measurements, including those that are not accurate enough and are statistically insignificant. To select reliable values of ${\cal B}$, we used the following criterion:
\be
\label{Eq.CritRealMF}
{\cal B}  > 2 \, \sigma_{\cal B}
\ee
For B stars, we investigated the dependence of the mean magnetic field for these stars on the projected rotational velocity $V\sin i$. We established the absence of any correlation between the mean field of B stars and their projected rotational velocity that is already known from the studies of other researchers, which is consistent with the hypothesis about a fossil origin of their magnetic fields.

\subsection{Magnetic Field Distribution Function for B Stars}
\label{ss.MFdistrFunct}

Analysis of the differential field distribution function $f({\cal B})$ (the magnetic field function) introduced by \cite{1997PAZh...23...47F} is of great importance in understanding the origin of stellar magnetic fields. The function $f({\cal B})$ is defined as 
\be
\label{Eq.f(B)determ}
N({\cal B},{\cal B}+\Delta {\cal B})\approx N f({\cal B})\Delta {\cal B} \, ,
\ee
where $N({\cal B},{\cal B}+\Delta {\cal B})$ is the number of stars in the interval of mean magnetic fields $({\cal B},{\cal B}+\Delta {\cal B})$, $N$ is
the total number of stars with measured ${\cal B}$. In this paper, we restrict ourselves only to B stars, because the number of O stars with measured magnetic fields is insufficient for statistical studies.

After the application of criterion (10), there were 130 objects in the list of stars with statistically significant fields. The distribution function $f({\cal B})$ constructed from the data of the catalog by Bychkov et al. (2009) is shown in Fig.~4. We chose the bins of mean magnetic fields in such a way that at least
eight stars fell within each bin. Only in the regions ${\cal B}<0.06\,$kG and ${\cal B}>5\,$kG did the number of stars turn out to be smaller than eight, because the number of stars with very small and very large magnetic field strengths was small.

The derived function $f({\cal B})$ for ${\cal B}\ge 400\,$G can be fitted by a power law:
\be
\label{Eq.meanBdistr}
f({\cal B})= A_0 \left(\frac{{\cal B}}{{\cal B}_0} \right)^{-\gamma} \, .
\ee
It turned out that in a wide range of ${\cal B}$ (0.40~--~12 kG), the distribution function $f({\cal B})$ could be described by a single expression (12) with parameters $A_0=0.33\pm 0.04$ and $\gamma=1.82\pm 0.07$, as shown in Fig.~4.

\cite{2000PreprintSAO..150} constructed the magnetic field function from a sample of 57 bright ($V<4^m.0$) magnetic main-sequence B3~--~F9 stars. These authors
fitted the magnetic field function by a power law. For ${\cal B}_s>4\,$kG, where ${\cal B}_s$ is the stellar surface field, which is approximately triple the value of ${\cal B}$ \cite{2000PreprintSAO..150}, the authors obtained $\gamma=2.2$, which is close to the value found here. In the range of magnetic fields 1~--~6 kG, these authors obtained $\gamma \approx 1$ and concluded that there was a break in the magnetic field function in the range $B_s= 3 - 5\,$kG. Our data are consistent with the conclusion about the existence of such a break (see Fig.~4). However, since the number of stars with measured magnetic fields in the above range is relatively small, the value of the parameter $\gamma$ in this range cannot be established reliably.

In their recent paper, \cite{2008AstBu..63..139R} presented a catalog of magnetic fields for CP stars, including 355 B and A stars. Since a substantial fraction of the objects listed in this catalog and considered here overlap, the magnetic field distribution function constructed from the data of the catalog by \cite{2008AstBu..63..139R} must be similar to that obtained here. Comparison of the histogram of ${\cal B}_{\mbox{\sz extr}}$ shown in Fig.~1 from the cited paper and our normalized distribution function shows that there is general agreement between the two distributions, although there are also deviations for ${\cal B}< 1\,$kG related to the greater detail of the distribution function in our paper in this range. Note that the exponential fit to the distribution of ${\cal B}_{\mbox{\sz extr}}$ used by \cite{2008AstBu..63..139R} does not seem optimal to us, because, in this case, the errors of the fit are considerably larger than those for the power-law fit (12).

The behavior of the function $f({\cal B})$ at relatively low values of ${\cal B}<400\,$G is of particular interest. The corresponding values of the function $f({\cal B})$ calculated using Eq.~(11) are indicated in Fig.~4 by the arrows. For such values of B, the behavior of $f({\cal B})$ does not follow the dependence (12).

At small mean magnetic field strengths ${\cal B} \lem 100\,$G, the values of the function $f({\cal B})$ are lower than those obtained from the fit (12) by more than an order of magnitude. Thus, we may conclude that the empirical magnetic field distribution function decreases sharply at ${\cal B}$ below a threshold value of ${\cal B}^{\mbox{\scriptsize thresh}} \approx 400\,$G.

The deviations of $f({\cal B})$ from the power law (12) may be related to the difficulty of detecting relatively weak magnetic fields. To estimate the effect of non-detection of weak magnetic fields, let us consider the following model.

Let a large number of magnetic field measurements for various stars be performed with a spectropolarimeter that measures the field with an accuracy $\sigma_{\cal B}$. Let us calculate the probability of detecting the weak magnetic field of a star with a mean magnetic field ${\cal B}$ for $n$ random field measurements.

Suppose that the model star for which the field is measured possesses a dipole magnetic field with a polar field strength ${\cal B}_p$ and $N_{tot}\gg 1$ random values of the inclination of the rotation axis $i$ and the angle $\beta$ and that the observations are performed at $n$ random rotation phases $\phi$. We will use the following field detection criterion: if the absolute value of the longitudinal field component $|B_l|$ is higher than $3\sigma_{\cal B}$ for $k\lem n$ random rotation phases at the time of field measurements, then we will assume the field to have been detected. Then,
\be
\label{Eq.ProbFindMF}
P(n,\sigma_{\cal B},{\cal B},k) = \frac{N_{\mbox{\sz detect}}}{N_{\mbox{\sz tot}}} \, ,
\ee

where $N_{\mbox{\sz detect}}$ is the number of measurements in which the field was detected according to the above criterion. Below, we will use $k=2$ adopted for many observational works and denote $P(n,\sigma_{\cal B},{\cal B},2) = P(n,\sigma_{\cal B},{\cal B})$. The probabilities $P(n,\sigma_{\cal B},{\cal B})$ calculated for $n=6$ are presented in Fig.~5. Note that for $n>6$ the field detection probabilities are almost constant.

Let us describe the procedure of reconstructing the real distribution function from the observed one and ascertain how the magnetic field distribution
would appear at a given measurement error $\sigma_{\cal B}$ if the probability of detecting weak fields was described by Eq. (13).

Since $P({\cal B},\sigma_{\cal B},n)\approx 1$ for ${\cal B}>400\,$G and at $\sigma_{\cal B}=100\,$G, which generally exceeds the errors of present-day magnetic field measurements, we will assume that a magnetic field with ${\cal B}>{\cal B}_{\mbox{\sz c}}=400\,$G was detected in all of the stars for which the corresponding measurements were made. Let ${\cal B}_{\mbox{\sz min}}$ be the minimum mean magnetic field that can still be determined at the present-day measurement accuracy. Based on the data from \cite{2007A&A...475.1053A}, \cite{2008ApJ...686.1269M}, and \cite{2008A&A...483..857S}, we can conclude that ${\cal B}_{\mbox{\sz min}}\approx 25\,$G.

We will assume the magnetic field distribution function in the entire range ${\cal B}>{\cal B}_{\mbox{\sz min}}$ to be described by the power law (12). The normalized distribution function will be 
\be
\label{Eq.MFFreal}
f^{\mbox{\sz real}}({\cal B})=A_* {\cal B}^{-\gamma} \, ,
\ee
where ${\cal B}_0=1\,$kG, i.e., ${\cal B}$ is measured in kG, and the factor $\ds A_*=(\gamma-1){\cal B}_{\mbox{\sz min}}^{\gamma-1}$ is defined by the normalization condition
$\ds 
\int_{{\cal B}_{\mbox{\sz min}}}^{\infty}\!\!\!f^{\mbox{\sz real}}({\cal B})d{\cal B}=1 \, .
$
Let the field detection probability be equal to the function $P({\cal B},\sigma_{\cal B},n)$ described above. The number of stars that will be detected in the interval $({\cal B},{\cal B}+\Delta {\cal B})$ at a given value of $\sigma_{\cal B}$ is then
\be
\label{Eq.fsigmaRelation}
N({\cal B},{\cal B}+\Delta {\cal B})\Delta {\cal B}= 
N_{\sigma_{\tiny{\cal B}}} f_{\sigma_{\tiny{\beta}}}({\cal B})\Delta{\cal B}= 
N_* P({\cal B},\sigma_{\cal B},n) f^{\mbox{\sz real}}({\cal B}) \Delta {\cal B} \, 
\ee
Here, $N_{\sigma_{B}}$ is the number of stars in which a magnetic field will be detected at a given value of $\sigma_{\mbox{\tiny B}}$, $N_*$ is
the total number of stars in the ensemble of magnetic stars with mean field strength distributed according to the law (14).

It follows from Eq. (15) that
\be
\label{Eq.fsigmafinal}
 f_{\sigma_B}({\cal B}) = P({\cal B},\sigma_{\cal B},n)  f^{\mbox{\sz real}}({\cal B}) \,
\frac{N_*}{N_{\sigma_B}}  = 
P({\cal B},\sigma_{\cal B},n)  f^{\mbox{\sz real}}({\cal B})\frac{1}{{\cal Q_{\sigma_B}}}\, .
\ee

The correction factor ${\cal Q_{\sigma_{B}}}$ allows for the fact that $N_{\sigma_B} < N_*$ at a nonzero $\sigma_{\cal B}$.

Based on Eq. (15), we can easily find that
\be
\label{Eq.Qcorr}
{\cal Q_{\sigma_B}} = \frac{N_{\sigma_B}}{N_*}  = \int\limits_{{\cal B}_{\mbox{\sz min}}}^{\infty}\!\! 
P({\cal B},\sigma_{\cal B},n) f^{\mbox{\sz real}}({\cal B}) d{\cal B}  \, .
\ee

The values of $f_{\sigma_B}({\cal B})$ calculated from Eq.~(16) are presented in Fig.~6. It can be said that $f_{\sigma_B}({\cal B})$ is the field distribution function that would be obtained during the observations of an ensemble of magnetic stars with mean fields distributed according to the law (14) when using an ideal spectropolarimeter that
measures the longitudinal field component for all of the observed stars with an accuracy $\sigma_{\cal B}$.

Nevertheless, the explanation of the cutoff in the magnetic field function for ${\cal B}<\,400\,$G may be incomplete, because the field measurement accuracy $\sigma_{\cal B}$ is currently fairly high, 20--100~G (\cite{2007A&A...475.1053A}; \cite{2008ApJ...686.1269M}; \cite{2008A&A...483..857S}). This allows magnetic fields to be also detected in the range 40--120~G. We see from Fig.~6 that a considerably larger number of stars with magnetic fields must be detected in this range at such values of $\sigma_{\cal B}$.

There exists an alternative explanation for the rapid decrease in $f({\cal B})$ in the range ${\cal B}<200-300\,$G.
\cite{2000mfcp.proc..149G} suggested that if the mean stellar magnetic field is below some threshold value of ${\cal B}^{\mbox{\scriptsize thresh}}$, then the field strength in the stellar atmosphere decreases almost to zero in a short time due to the processes of meridional circulation.

\cite{2007A&A...475.1053A} investigated a sample of 28 magnetic Ap/Bp stars; for 24 program stars, they fitted the phase dependence of the measured longitudinal field $B_l$ in the model of an oblique rotating dipole and obtained the polar field strengths $B_p$. A histogram of the number of stars in bins $\Delta \log {B_p}=0.2\,$dex was constructed from the values of $B_p$ obtained.

The number of stars with $B_p<B_p^{\mbox{\sz thresh}}\approx 300\,$G was found to be very small. Based on this fact, the authors suggested that stable configurations of the global stellar magnetic field exist at $B_p>B_p^{\mbox{\sz thresh}}$, while the global magnetic field is destroyed on the Alfven time scale through magnetic field instabilities at $B_p< B_p^{\mbox{\sz thresh}}$. According to \cite{1999A&A...349..189S}, \cite{2004IAUS..215..356S}, the most important type of instability is the pinch one.

Note that the threshold value of $B^{\mbox{\sz thresh}}_p \approx 300\,$ obtained by \cite{2007A&A...475.1053A}
corresponds to thresh $B^{\mbox{\scriptsize thresh}}\approx B_p^{\mbox{\scriptsize thresh}}/5 = 60\,$G, which is a factor of 6--7 lower than our threshold value of $B^{\mbox{\scriptsize thresh}} \approx 400\,$G. Such a significant discrepancy may stem from the fact that the number of objects with $B_p$ measured by \cite{2007A&A...475.1053A} is small. In addition, it should be noted that only 25$\%$ of these objects are B stars, while the remaining stars are A and F ones, for which the statistical properties of the ensemble of magnetic fields can differ significantly from those typical of B stars.

In addition, analysis of Fig.~6 in \cite{2007A&A...475.1053A} shows that there may exist another sharp decrease in the number of stars at $\ln(B_p)\approx 1500\,$G, corresponding to the mean magnetic field ${\cal B}=300\,$G, which is close to the threshold magnetic field obtained here.

\subsection{Dependence of the Magnetic Field Strength on the Stellar Main-Sequence Lifetime}

The question of how the magnetic field of a star changes during its stay on the main sequence is very important. The mean magnetic field ${\cal B}$ is plotted against the stellar main-sequence lifetime for B4--B9 star in Fig.~7. We took the relative stellar mainsequence lifetimes $\tau$ from \cite{2006A&A...450..763K} and \cite{2006MNRAS.372.1804K} and the mean magnetic field strengths ${\cal B}$ from the catalog by \cite{2009MNRAS.394.1338B}.

We see from Fig.~7 that ${\cal B}(\tau)$ decrease regularly with increasing relative stellar main-sequence life-time $\tau$ in agreement with the conclusion reached by \cite{2006A&A...450..763K}.

\cite{2008A&A...481..465L} analyzed the dependence of the rms magnetic field ${\cal B}$ and magnetic flux ${\cal F}={\cal B}R^2$ for A and B stars on their main-sequence lifetime. For early A and late B stars with masses 3--5$M_{\odot}$ , the variations in both ${\cal B}$ and ${\cal F}$ were found to be small, except for the first 15--20$\%$ of the stellar main-sequence lifetime. According to \cite{2008A&A...481..465L}, the mean magnetic field and magnetic flux of these stars in the range $\tau\in(0.0-0.2)$ decrease by a factor of 3--4 and then remain almost constant.

Since the variations in the radius of a star during its main-sequence evolution are insignificant, it can be concluded that the variations in mean stellar magnetic field presented in Fig.~7 and the variations in stellar magnetic flux may be related to the dissipation of weak magnetic fields noted above.

To explain the observed magnetic field function $f({\cal B})$, particularly for ${\cal B} < 400\,$G, it is necessary
to consider its evolution by taking into account the decrease in rms magnetic field ${\cal B}$ with increasing $\tau$. Such a study is being planned.

\section{CONCLUSIONS}
The following conclusions can be drawn from our analysis.

\begin{itemize}

\item (1) We reached a tentative conclusion about a possible significant (by more than a factor of 3) increase in the magnetic fields of OB stars averaged over the spectral subtypes when passing from O stars to B stars.

\item (2) No correlation was found between the mean magnetic field strength ${\cal B}$ and projected rotational velocity of OB stars, which is consistent with the hypothesis about a fossil origin of the magnetic field in these stars.

\item (3) We constructed the mean magnetic field distribution function for B stars, $F({\cal B})$, that has a power-law dependence on ${\cal B}$ with an exponent of $\approx -1.82$. We found a sharp decrease in the function $F({\cal B})$ for B < 400 G, which may be related to rapid dissipation of weak stellar surface magnetic fields.

\item (4) We confirmed the conclusion by \cite{2008A&A...481..465L} that over the main-sequence lifetime of a B star, its mean magnetic field can decrease by a factor of 5--7.
\end{itemize}

\section{ACKNOWLEDGMENTS}
This study was supported by the Program of the President of the Russian Federation for support of leading scientific schools (NSh-1318.2008.2).


\newpage
\begin{figure}
\centering
\includegraphics[width=16 cm]{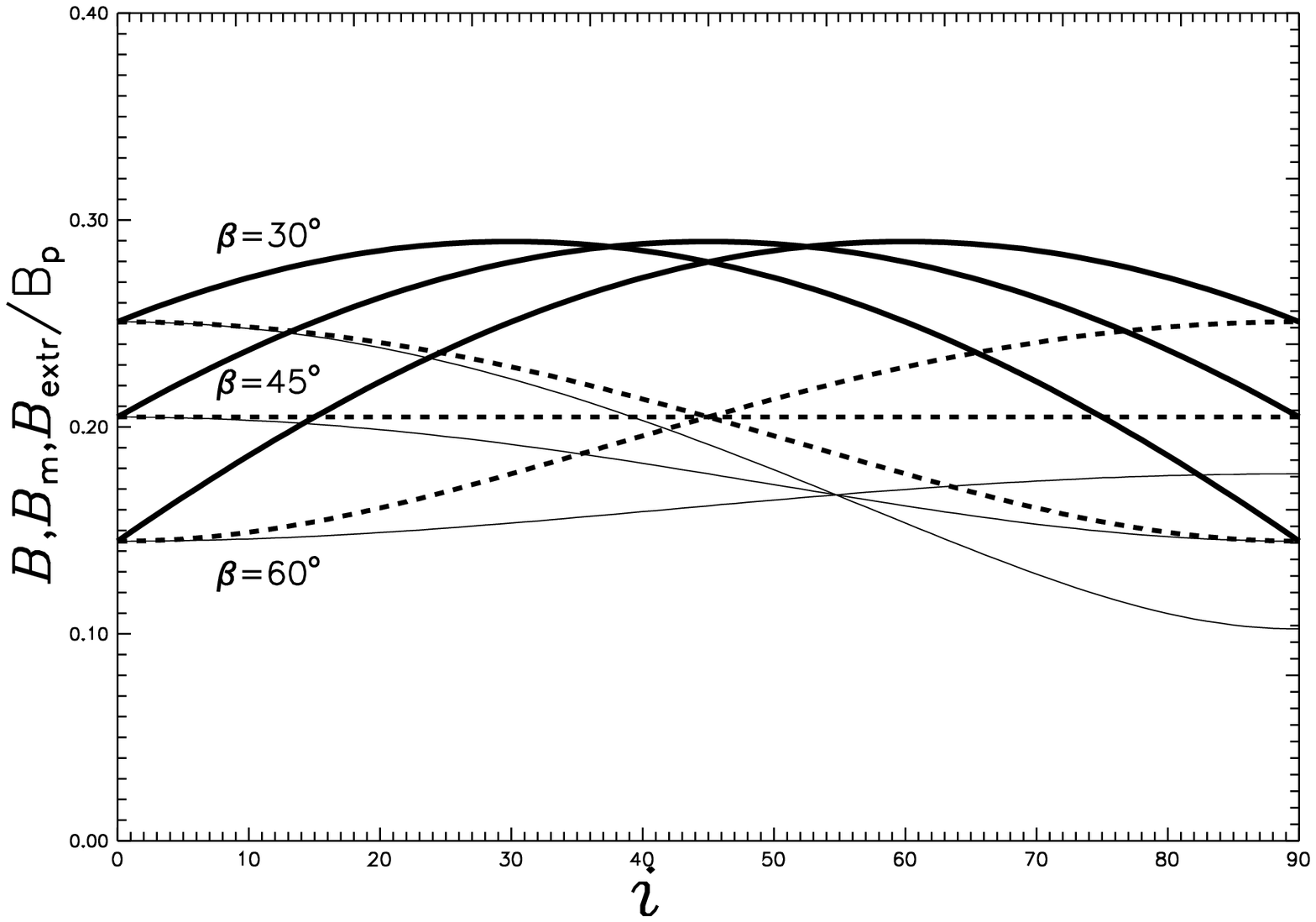}
\caption{Rotation-phase-averaged strenght of the mean longitudinal magnetic field versus inclination of rotation axis $\ds i$:
the thin solid, dashed, and thick solid lines represent ${\cal B}/B_p$, ${\cal B}_m/B_p$, ${\cal B}_{\mbox{\sz extr}}/B_p$ respectively. The angles $\beta$ are marked.}
\end{figure}

\begin{figure}
\centering
\includegraphics[width=16 cm]{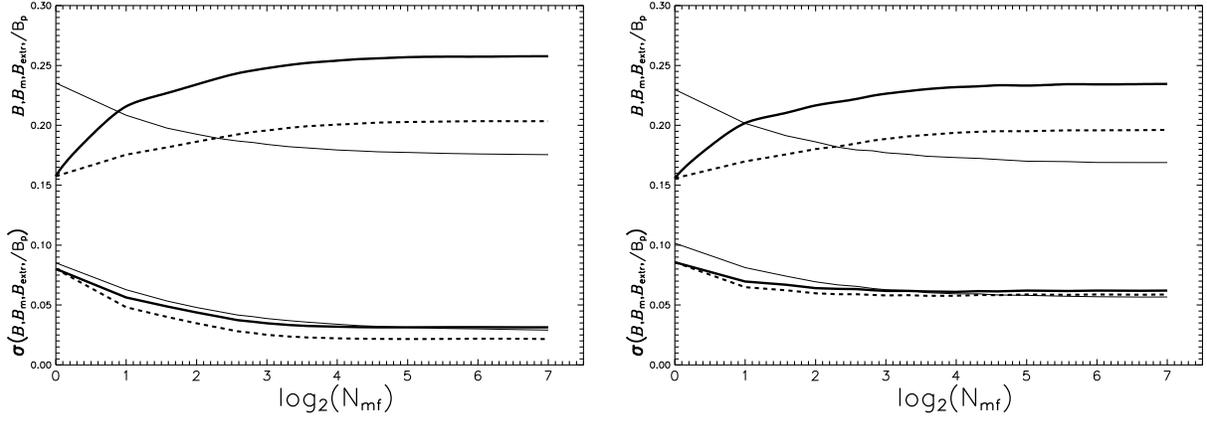}
\caption{The $\bf upper$ $\bf curves$ represent the rotation-phase-averaged strength of the mean longitudinal magnetic field as a function of the logarithm of number of field ''measurements'' for random values of the angles $\ds i$ and $\beta$. The $\bf lower$ $\bf curves$ represent the corresponding standard deviation. The angle $\ds i$ in both figures varies over the range 
$0-90^{\circ}$; the angle $\beta$ lies within the ranges $30-60^{\circ}$ (a) and $0-90^{\circ}$ (b). The notation is the same as that in Fig.~1.}
\end{figure}

\begin{figure}
\centering
\includegraphics[width=16 cm]{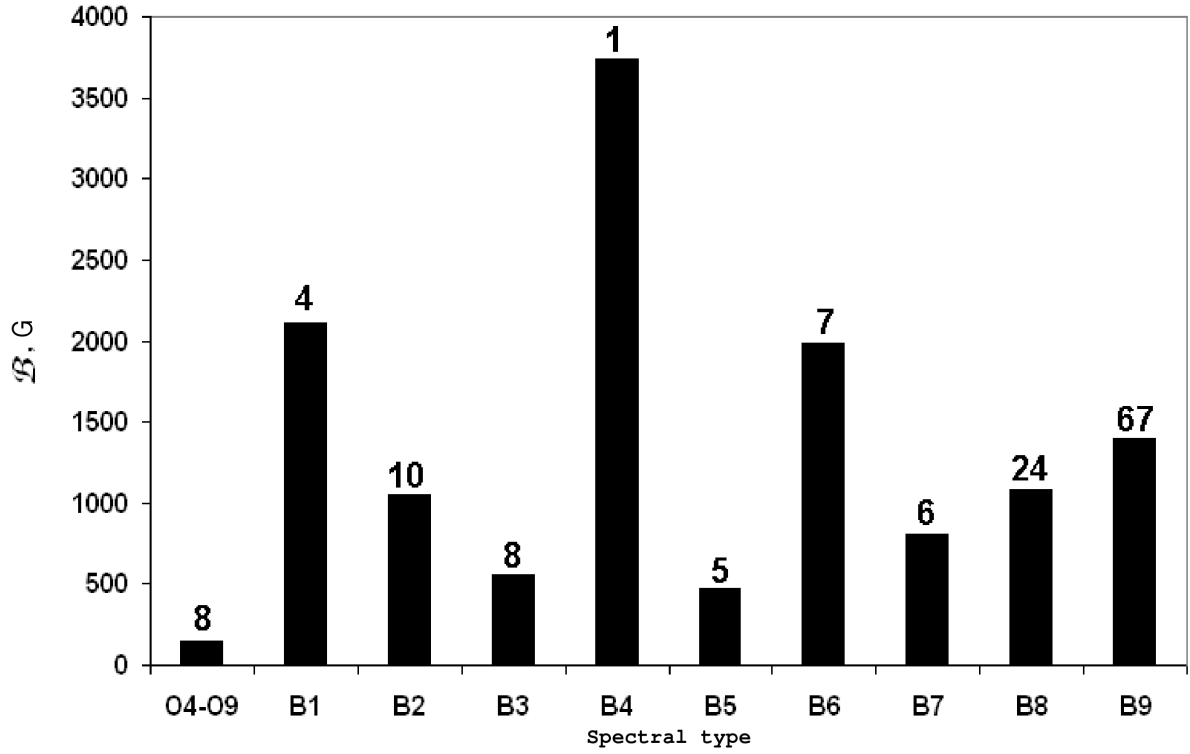}
\caption{Distribution function of the spectral-type-averaged magnetic field for OB stars. The numbers mark the number of stars with measured magnetic fields satisfying the criterion (10).}
\end{figure}

\begin{figure}
\centering
\includegraphics[width=16 cm]{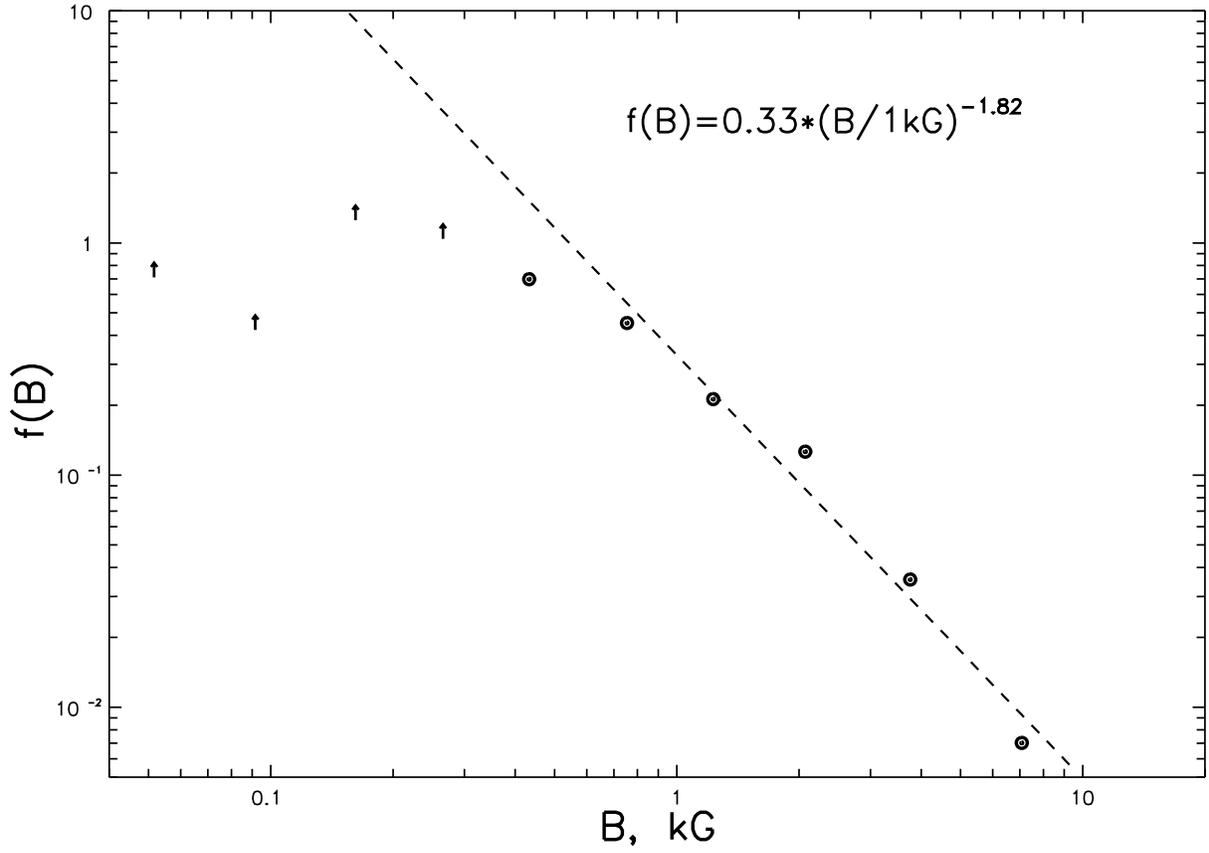}
\caption{Magnetic field distribution function for B stars: the points represent the mean values of $f({\cal B})$ for  ${\cal B} \ge 400\,$ G; the vertical arrows represent $f({\cal B})$ for ${\cal B} < 400\,$ G.}
\end{figure}

\begin{figure}
\centering
\includegraphics[width=16 cm]{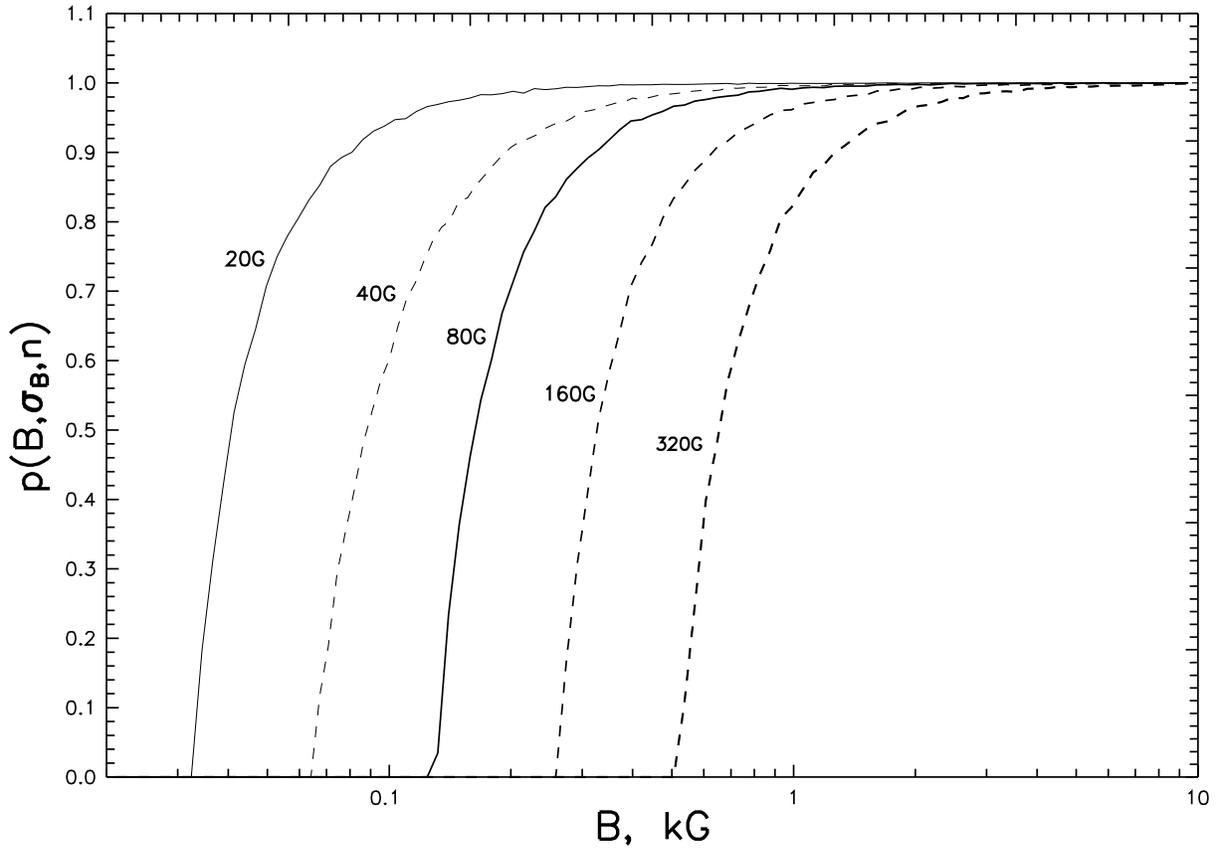}
\caption{Probability $P(n,\sigma_{\cal B},{\cal B})$ of detecting the magnetic field of a star with a mean magnetic field ${\cal B}$ at $n=6$. The values of $\sigma_{\cal B}$ are indicated near the corresponding curves.}
\end{figure}

\begin{figure}
\centering
\includegraphics[width=16 cm]{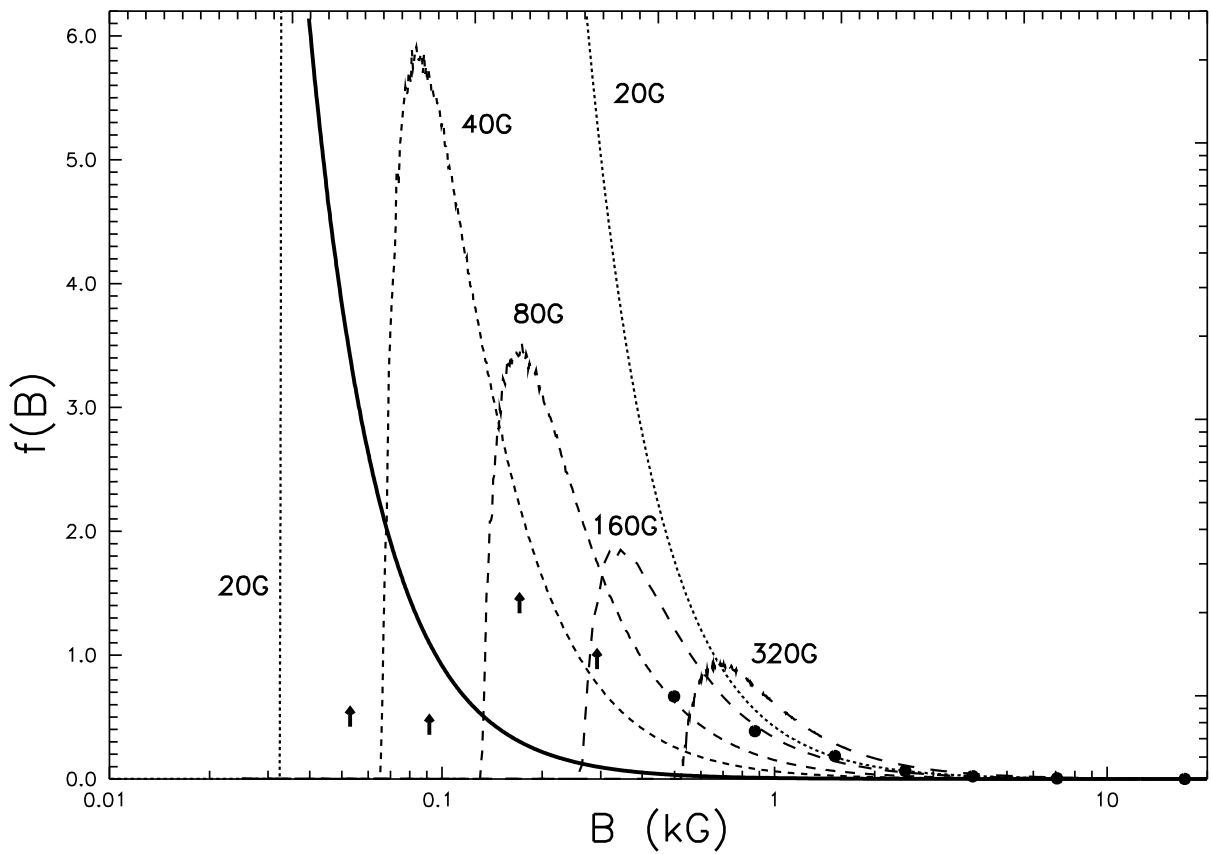}
\caption{The magnetic field distribution function calculated by assuming that the power law (14) is valid in the entire range of measured magnetic filds (solid line). The magnetic field distribution functions for OB stars corrected for the possible nondetection of weak magnetic fields at a measurement accuracy $\sigma_{\cal B}$ (dashed lines). The values of $\sigma_{\cal B}$ are indicated near the corresponding curves. The notation is the same as that in Fig. 4.}
\end{figure}

\begin{figure}
\centering
\includegraphics[width=16 cm]{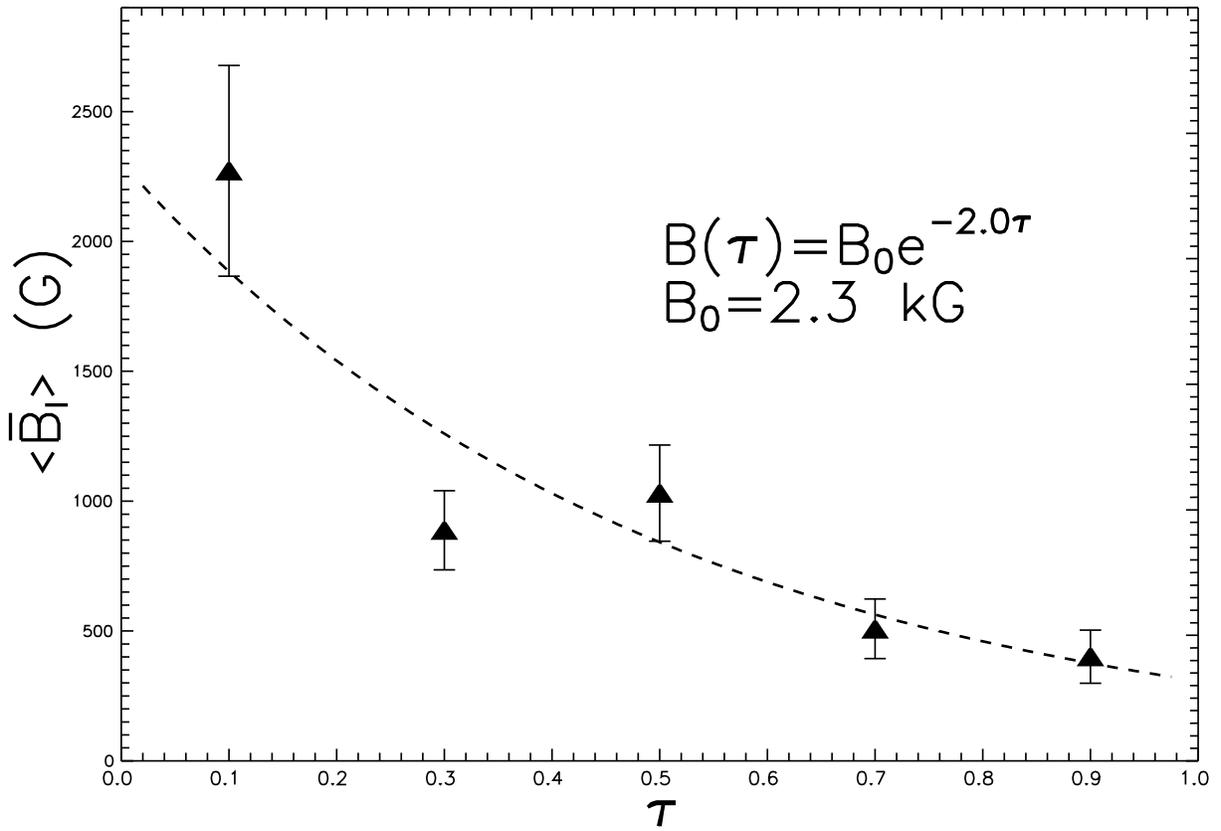}
\caption{Mean magnetic field strength ${\cal B}$ for B stars versus relative stellar main-sequence lifetime $\tau$: the triangles represent ${\cal B}$ averaged over the bins $\Delta\tau=0.2$; the curve indicates an exponential fit to the dependence ${\cal B}(\tau)$.}
\end{figure}
\end{document}